\documentclass[amsmath,amsfonts,amssymb,twocolumn,nofootinbib]{revtex4-1}
\usepackage{geometry}
\usepackage{graphicx}
\usepackage{amssymb, amsmath}
\usepackage{epstopdf}
\usepackage{subfigure}
\usepackage[table]{xcolor}
\usepackage{algcompatible}
\usepackage[floatrow]{trivfloat}
\usepackage{tikz}

\graphicspath{{Figures/}}

\usetikzlibrary{intersections, backgrounds}

\trivfloat{algorithm}

\definecolor{highlight}{RGB}{180, 31, 180}
\definecolor{gray80}{gray}{0.8}

\definecolor{comment}{gray}{0.50}

\begin{document}

\title{Adiabatic Monte Carlo}
\author{Michael Betancourt}
\affiliation{Department of Statistics, University of Warwick, Coventry CV4 7AL, UK}
\email{betanalpha@gmail.com}

\date{\today} 

\begin{abstract}

A common strategy for inference in complex models is the relaxation of a simple
model into the more complex target model, for example the prior into the posterior 
in Bayesian inference.  Existing approaches that attempt to generate such 
transformations, however, are fragile and can be difficult to implement effectively 
in practice.  Leveraging the geometry of equilibrium thermodynamics, I introduce 
a principled and robust approach to deforming measures that presents a powerful 
new tool for inference.

\end{abstract}

\maketitle

Bayesian inference provides an elegant approach to inference by summarizing
information about a system in a probabilistic model and formalizing inferential 
queries as expectations with respect to that model.  Although conceptually 
straightforward, this approach was long limited in practice due to the computational 
burden of computing these expectations, especially for the high-dimensional 
distributions of practical interest.

\begin{figure*}
\centering
\subfigure[]{ \includegraphics[width=2.8in]{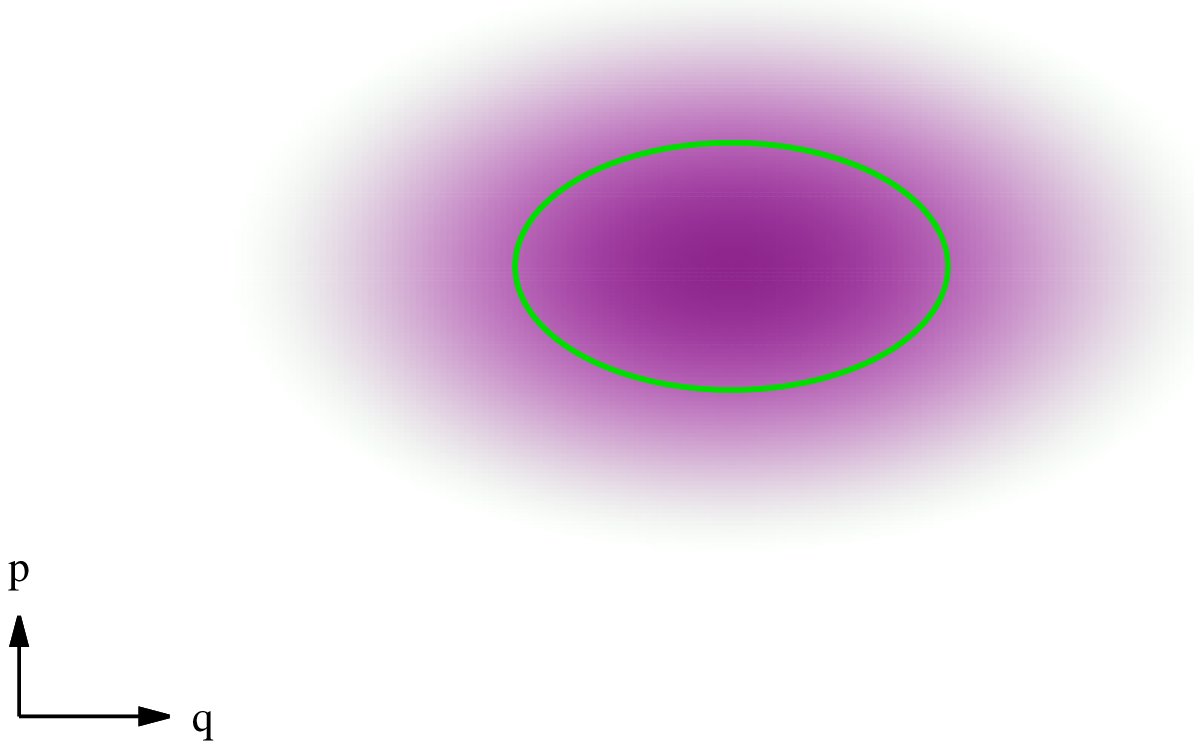}}
\subfigure[]{ \includegraphics[width=2.8in]{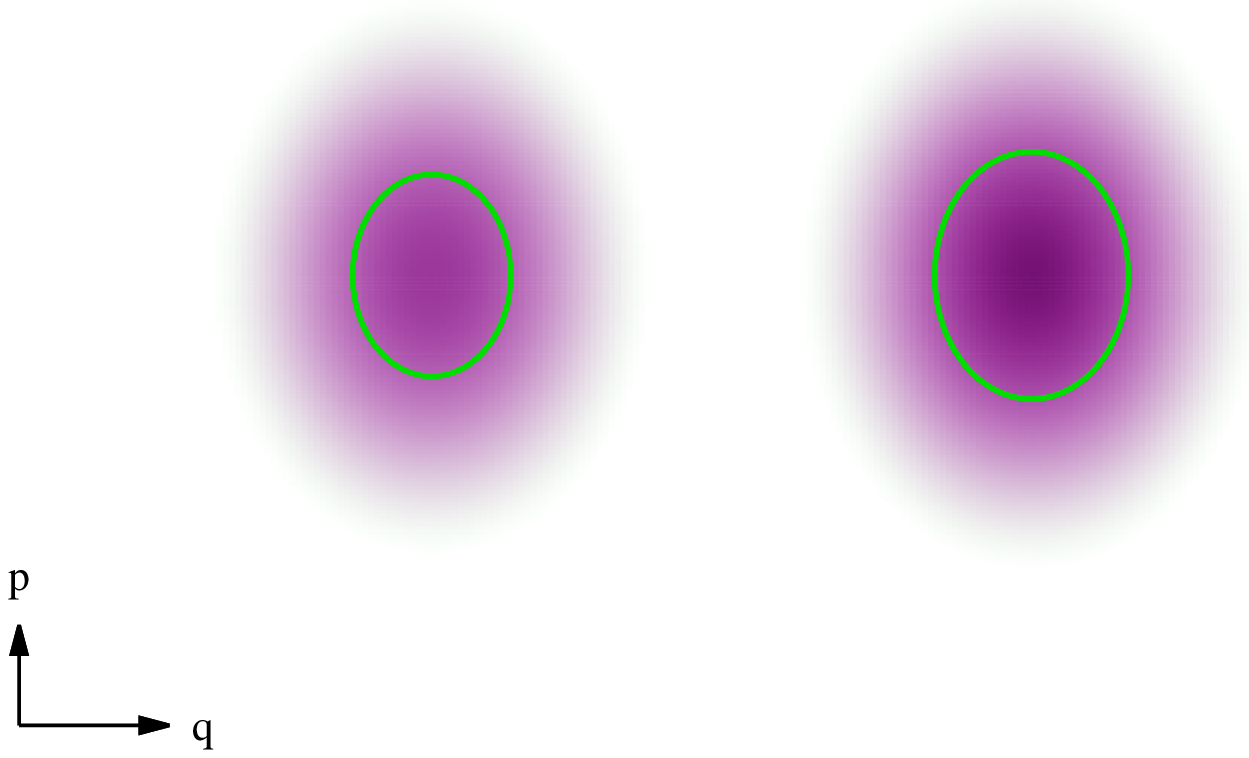}}
\caption{Hamiltonian Monte Carlo generates trajectories that efficiently explore level 
sets of constant probability, and hence the target distribution itself.  (a) The
level sets of unimodal target distributions are simply-connected and can be explored
with a single trajectory using only local information of the target distribution. (b) On the
other hand, the level sets of multimodal target distributions are disconnected, requiring 
multiple trajectories and global information of the target distribution to ensure 
comprehensive exploration.  Deforming the multimodal target distribution into a
unimodal one, however, warps the disconnected level sets into connected ones
and dramatically eases sampling.}
\label{fig:level_sets}
\end{figure*}

Markov Chain Monte Carlo \cite{RobertEtAl:1999, BrooksEtAl:2011}, 
revolutionized the practice of Bayesian inference by using localized information 
from the model to estimate expectations.  Provided that the model is itself localized, 
Markov Chain Monte Carlo then yields computationally efficient estimates.
When the model features more complex global structure such as multimodality, 
however, those estimates become much less satisfactory.  This is particularly evident 
in Hamiltonian Monte Carlo \cite{DuaneEtAl:1987, Neal:2011, BetancourtEtAl:2014} 
where multimodality manifests as a nontrivial topological structure 
(Figure \ref{fig:level_sets}).

One approach to improving the validity of Markov Chain Monte Carlo estimates
in these situations is to deform the complicated model into a simpler, more
well-behaved model.  In particular, a measure-preserving bijection will
map easily-generated samples from the simple distribution into the desired 
samples from the complex distribution.  This approach has motivated a variety of 
statistical algorithms in the literature that, while successful in some applications,
are ultimately limited by their own construction.

In this paper I present a principled means of constructing measure-preserving
deformations of a simple distribution into an arbitrarily complicated one by leveraging
the geometry of equilibrium thermodynamic processes, namely contact manifolds.  
After discussing the limitations of existing approaches I introduce contact Hamiltonian 
flows, discuss their application to probabilistic systems, and demonstrate their utility 
as a Markovian transition on a simple example.

\section{Thermodynamic Algorithms}

An immediate strategy for deforming a complex distribution into a simpler one is
the moderation of the density between the two distributions.  

Consider a topological sample space, $Q$, its usual Borel $\sigma$-algebra, 
$\mathcal{B} \! \left( Q \right)$, and a potentially-complex target 
distribution, $\pi$.  Assuming absolute continuity, we can construct
the target distribution from a unimodal and otherwise well-behaved 
base measure, $\pi_{B}$, and a density incorporating any complicated, 
possibly multimodal structure,
\begin{align*}
\pi 
&= \frac{ \mathrm{d} \pi }{ \mathrm{d} \pi_{B} } \pi_{B} \\
&\equiv e^{ - \Delta V } \pi_{B}.
\end{align*}
We can then generate a continuum of distributions between $\pi_{B}$ and $\pi$
by exponentiating the density,
\begin{align*}
\pi_{\beta}
&= 
\frac{1}{ Z \! \left( \beta \right) }
\left( \frac{ \mathrm{d} \pi }{ \mathrm{d} \pi_{B} } \right)^{\beta} \pi_{B}
\\
&=
\frac{ e^{- \beta \, \Delta V } \pi_{B} }
{ Z \! \left( \beta \right) },
\end{align*}
where
\begin{equation*}
Z \! \left( \beta \right) = \int_{Q} \mathrm{d} \pi_{B} \, e^{- \beta \, \Delta V }.
\end{equation*}
The desired deformation now takes the form of a bijection
between any two intermediate distributions,
\begin{align*}
f :& \, Q \times \mathbb{R} \rightarrow Q \times \mathbb{R}
\\
& \left(q, \beta\right) \mapsto \left( q', \beta' \right),
\end{align*}
which ideally maintains equilibrium,
\begin{equation} \label{measure_preserving_bijection}
f_{*} \pi_{\beta} = \pi_{\beta'}.
\end{equation}
Methods traversing this spectrum of distributions are often analogized with 
thermodynamics, where $\beta$ takes the role of an inverse temperature and 
$Z \! \left( \beta \right)$ the partition function.  

In \textit{simulated annealing} \cite{KirkpatrickEtAl:1983, Cerny:1985, Neal:1993} a 
deformation is generated by deterministically pushing $\beta$ along a rigid partition known 
as a \textit{schedule}, with the state stochastically evolved in between temperature updates 
using a Markov chain targeting the current $\pi_{\beta}$.  The performance of simulated 
annealing depends crucially on the sensitivity of $\pi_{\beta}$ to $\beta$ -- because the 
temperature is changed with the state held constant there is no guarantee that the state 
will remain in equilibrium with respect to the new $\pi_{\beta}$.

\textit{Simulated tempering} \cite{MarinariEtAl:1992, Neal:1993} appeals to the same
rigid partition but ensures equilibrium by applying a Metropolis correction to each 
move along the partition.  Formally, this generates a Markov chain with transitions 
proposing the exchange of states at temperatures $\beta$ and $\beta \pm \delta \beta$ 
with acceptance probability
\begin{equation*}
p \! \left( \mathrm{accept} \right) = \min \! \left( 1, 
\frac{ \pi_{\beta} \! \left( q_{1} \right) }{ \pi_{\beta \pm \delta \beta} \! \left( q_{1} \right) }
\frac{ \pi_{\beta \pm \delta \beta} \! \left( q_{2} \right) }{ \pi_{\beta} \! \left( q_{2} \right) }
\right).
\end{equation*}
The cost of maintaining equilibrium is that the random exploration of the temperature
partition proceeds only slowly, especially when $\pi_{\beta}$ rapidly varies with $\beta$.

Ultimately both approaches are limited by their dependence on a rigid partition of 
temperatures.  When $\pi_{\beta}$ is highly-sensitive to $\beta$ the probability mass 
rapidly changes with temperature, frustrating equilibrization in simulated annealing and 
intensifying random walk behavior in simulated tempering.  Formally, the sensitivity 
can be quantified with the Kullback-Leibler divergence between any neighboring 
distributions,
\begin{align*}
\mathrm{KL} \! \left( \pi_{\beta} || \pi_{\beta + \delta \beta} \right)
&=
- \int_{Q} \mathrm{d} \pi_{\beta} 
\log \frac{ \mathrm{d} \pi_{\beta + \delta \beta} }{ \mathrm{d} \pi_{\beta} }
\\
&=
- \int_{Q} \mathrm{d} \pi_{\beta} \log e^{ - \delta \beta \, \Delta V }
\\
& \quad - 
\int_{Q} \mathrm{d} \pi_{\beta} 
\log \frac{ Z \! \left( \beta \right) }{ Z \! \left( \beta  + \delta \beta \right) }
\\
&= -
\delta \beta \int_{Q} \mathrm{d} \pi_{\beta} \left( - \Delta V \right)
\\
&\quad -
\log \frac{ Z \! \left( \beta \right) }{ Z \! \left( \beta + \delta \beta \right) }
\int_{Q} \mathrm{d} \pi_{\beta} 
\\
&=
- \delta \beta \frac{1}{ Z \! \left( \beta \right) } \frac{ \partial Z }{ \partial \beta }
 \! \left( \beta \right)
 \\
& \quad - \log \frac{ Z \! \left( \beta \right) }{ Z \! \left( \beta + \delta \beta \right) }.
\end{align*}

Without any dynamic adaptation, the optimal performance of both algorithms is 
achieved when the deformation is constant across the partition,
\begin{equation*}
\mathrm{KL} \! \left( \pi_{\beta} || \pi_{\beta + \delta \beta} \right) \approx \mathrm{const}.
\end{equation*}
Because the partition function is rarely known a priori, however, determining an 
effective gradation is usually impossible and the algorithms must instead rely on 
adaptation schemes that themselves are sensitive to the details of the target 
distribution and its evolution along the partition.

\section{Adiabatic Monte Carlo}

This fragility of simulated annealing and simulated tempering to the temperature 
schedule arises because proposed moves across the partition do not themselves 
preserve the intermediate distributions as in \eqref{measure_preserving_bijection}.  
Truly measure-preserving processes, however, arise naturally in the thermodynamic 
analogy as \textit{adiabatic processes}, which mathematically correspond
to special flows on \textit{contact manifolds} \cite{Geiges:2008, Lee:2013}.  By mapping 
a given probability space into a contact manifold we can canonically construct these 
flows, both in theory and in practice, which are capable of exactly mapping samples 
from $\pi_{B}$ into $\pi$.

\subsection{Contact Hamiltonian Flows and Adiabatic Processes}

A contact manifold is a $\left( 2n + 1 \right)$-dimensional manifold, $\mathcal{R}$, endowed 
with a \textit{contact form}, $\alpha$, satisfying
\begin{equation*}
\Omega_{C} = \alpha \wedge \left( \mathrm{d} \alpha \right)^{n} \neq 0;
\end{equation*}
because of this non-degeneracy condition $\Omega_{C}$ serves as a canonical
volume form and orients the manifold.  A given contact form and a \textit{contact 
Hamiltonian},  $H_{C} : \mathcal{R} \rightarrow \mathbb{R}$, uniquely identify a 
\textit{contact vector field} by
\begin{align*}
\alpha \! \left( \vec{X}_{H_{C}} \right) &= H_{C}
\\
\left. \mathrm{d} \alpha \! \left( \vec{X}_{H_{C}}, \cdot \right) \right|_{\xi}
&= 
\left. - \mathrm{d} H_{C} \right|_{\xi},
\end{align*}
where $\xi$ is the \textit{contact structure}, 
\begin{equation*}
\xi = \left\{ v \in T \mathcal{R} :  \alpha \! \left( v \right) = 0 \right\}.  
\end{equation*}

Locally any contact manifold decomposes into the product of a symplectic manifold and 
$\mathbb{R}$, yielding the canonical coordinates $\left( q^{i}, p_{i}, \gamma \right)$.
In these canonical coordinates the contact form becomes
\begin{equation*}
\alpha = \mathrm{d} \gamma + \theta,
\end{equation*}
where $\theta$ is the local primitive of the symplectic form, 
$\mathrm{d} \theta = - \Omega$. Any contact vector field then factors into three 
components,
\begin{align*}
\vec{X}_{H_{C}} =&
+\left( H_{C} - p_{i} \frac{ \partial H_{C} }{ \partial p_{i} } \right) 
\frac{\partial}{ \partial \gamma} 
\\
& + \left( 
\frac{ \partial H_{C} }{ \partial p_{i} } \frac{ \partial }{ \partial q^{i} }
- \frac{ \partial H_{C} }{ \partial q^{i} } \frac{ \partial }{ \partial p_{i} }
\right)
\\
& + \;\;\, \frac{\partial H_{C}}{ \partial \gamma} p_{i} \frac{ \partial }{ \partial p_{i} }:
\end{align*}
the first term is a \textit{Reeb vector field} generating a change in the contact coordinate,
$\gamma$; the second term is a symplectic vector field convolving the symplectic 
coordinates; and the final term is a \textit{Liouville vector field} that scales the $p$.

Unlike a Hamiltonian flow on a symplectic manifold, a contact Hamiltonian flow does 
not foliate the contact manifold.  In fact the largest integrable submanifolds consistent 
with a given contact structure,
\begin{equation*}
\mathcal{S} \subset \mathcal{R}, T \mathcal{S} \subset \xi,
\end{equation*}
are the $n$-dimensional \textit{Legendrean submanifolds}, and only the flowout of 
$H_{C}^{-1} \! \left( 0 \right)$ is constrained to such a submanifold.  Thermodynamically,
the image of $H_{C}^{-1} \! \left( 0 \right)$ along a corresponding contact Hamiltonian 
flow is exactly an adiabatic process \cite{Mrugala:1978}.

The statistical utility of adiabatic processes lies in the fact that they preserve both
the contact Hamiltonian,
\begin{align*}
\left. \mathcal{L}_{\vec{X}_{H_{C}}} H_{C}  \right|_{H_{C}^{-1} \left( 0 \right)}
&= 
\left. \mathrm{d}H \! \left( \vec{X}_{H} \right) \right|_{H_{C}^{-1} \left( 0 \right)} 
\\
&= \left. H_{C} \frac{ \partial H_{C} }{ \partial \gamma } \right|_{H_{C}^{-1} \left( 0 \right)}
\\
&= 0,
\end{align*}
and the contact form
\begin{align*}
\left. \mathcal{L}_{\vec{X}_{H}} \alpha \right|_{H_{C}^{-1} \left( 0 \right)} 
&=
\left. \mathcal{L}_{\vec{X}_{H}} \alpha \right|_{\xi}
\\
&= 
\left. \left( 
\mathrm{d} \alpha \! \left( \vec{X}_{H}, \cdot \right) 
+ \mathrm{d} \left( \alpha \! \left( \vec{X}_{H} \right) \right) 
\right) \right|_{\xi} 
\\
&= 
\left. \left( 
\mathrm{d} \alpha \! \left( \vec{X}_{H}, \cdot \right) + \mathrm{d} H
\right) \right|_{\xi} 
\\
&= 0.
\end{align*}

Consequently adiabatic processes also preserve the canonical measure,
$\pi_{H_{C}} = e^{ - H_{C} } \Omega_{C}$,
\begin{equation*}
\left. \mathcal{L}_{\vec{X}_{H_{C}}} 
\left( \pi_{H_{C}} \right) \right|_{H_{C}^{-1} \left( 0 \right)}  
= 0,
\end{equation*}
or
\begin{equation*}
\left. \left( \phi^{H_{C}}_{t} \right)_{*} 
\left( \pi_{H_{C}} \right) \right|_{H_{C}^{-1} \left( 0 \right)}
= \pi_{H_{C}}.
\end{equation*}
Recognizing heat in thermodynamics as probability, this measure-preservation 
corresponds to the property that adiabatic processes exchange no heat with 
the environment.

One concern with adiabatic processes is that, unlike their symplectic 
counterparts, contact Hamiltonian flows may have fixed points which 
prevent the corresponding adiabatic process from being a bijection, 
and hence preserving the canonical measure, for all $t$.

\subsection{Constructing Contact Hamiltonian Systems}

For these measure-preserving flows to be the basis of a useful statistical
algortihm, we first require a canonical way of manipulating a given probabilistic 
system into a contact Hamiltonian system.  Following the geometric construction 
of Monte Carlo \cite{BetancourtEtAl:2014}, we do this by first mapping the
probabilistic system into a Hamiltonian system which we then \textit{contactize} 
into a contact Hamiltonian system.

In order to construct a Hamiltonian system we first lift the base measure, 
$\pi_{B}$, to a measure on the cotangent bundle of the sample space, 
$\varpi: T^{*} Q \rightarrow Q$, with the choice of a disintegration, $\xi$,
\begin{equation*}
\pi_{H} = \varpi^{*} \pi_{B} \wedge \xi = e^{-H} \Omega,
\end{equation*}
where
\begin{equation*}
H = - \log \frac{ \mathrm{d} \pi_{H} }{ \mathrm{d} \Omega }.
\end{equation*}
In canonical coordinates, $\left(q^{i}, p_{i} \right)$, we have the decompositions
\begin{align*}
\pi_{B} &= e^{-V_{B}} \mathrm{d}^{n} q
\\
\xi &= e^{-T} \mathrm{d}^{n} p + \text{horizontal} \; n\text{-forms},
\end{align*}
in which case the Hamiltonian becomes
\begin{equation*}
H = T + V_{B}.
\end{equation*}

We can now contactize the cotangent bundle by affixing a contact coordinate,
$\gamma \in \mathbb{R}$,
\begin{equation*}
\mathcal{R} = T^{*} Q \times \mathbb{R},
\end{equation*}
with the corresponding contact form $\alpha = \mathrm{d} \gamma + \theta$, 
where $\theta$ is the tautological one-form on the cotangent bundle.  Finally, 
we lift $\pi_{H}$ to $\mathcal{R}$ by introducing the density,
$\mathrm{d} \pi / \mathrm{d} \pi_{B}$, and some constant, $H_{0}$,
\begin{align*}
\pi_{H_{C}} 
&= 
\frac{1}{Z \! \left( \beta ( \gamma ) \right) }
\left( \frac{ \mathrm{d} \pi }{ \mathrm{d} \pi_{B} } \right)^{ \beta \left( \gamma \right)} 
e^{-H_{0}} \alpha \wedge \pi_{H} 
\\
&= e^{-H_{C}} \Omega_{C},
\end{align*}
where
$H_{C}$ is the resulting contact Hamiltonian,
\begin{equation*}
H_{C} = 
T + V_{B} + \beta \! \left( \gamma \right) \, \Delta V + 
\log Z \! \left( \beta (\gamma ) \right) + H_{0}.
\end{equation*}
with $Z \! \left( \beta ( \gamma ) \right)$ the partition function,
\begin{equation*}
Z \! \left( \beta ( \gamma ) \right) = 
\int_{T^{*} Q} \left. \left( e^{- \left( T + V_{B} + \beta ( \gamma ) \, \Delta V + H_{0} \right)} 
\Omega_{C} \right) \right|_{\beta (\gamma) }.
\end{equation*}
In practice $H_{0}$ is chosen to ensure that the initial point lies in the
zero level set, $H^{-1}_{C} \! \left( 0 \right)$.

Now given a particular value of $\gamma$, the distribution $\pi_{H_{C}}$ 
restricts to a canonical distribution on the cotangent bundle,
\begin{align*}
\pi_{H_{\beta ( \gamma )}} 
&= \left. \pi_{H_{C}} \right|_{\beta (\gamma)} 
\\
&= e^{ - \left. H_{C} \right|_{\beta (\gamma)}  } \Omega
\\
&\equiv e^{ H_{\beta (\gamma)} } \Omega,
\end{align*}
which then projects down to a distribution on the target sample space,
\begin{align*}
\pi_{\beta (\gamma) }
&= \varpi^{*} \pi_{H_{\beta ( \gamma )}}
\\
&= 
\frac{1}{Z \! \left( \beta ( \gamma ) \right)}
\left( \frac{ \mathrm{d} \pi }{ \mathrm{d} \pi_{B} } \right)^{ \beta \left( \gamma \right)} 
\pi_{B}.
\end{align*}
In order to map $\gamma \in \mathbb{R}$ into the interval $\beta \in \left[ 0, 1 \right]$, 
I will assume the relationship
\begin{equation} \label{eqn:logistic}
\beta \! \left( \gamma \right) = \frac{1}{1 + e^{-\gamma}}
\end{equation}
through the rest of the paper.

Adiabatic processes are then generated by contact Hamiltonian flowout of 
$H_{C}^{-1} \! \left( 0 \right)$,
\begin{align*}
\left. \vec{X}_{H_{C}} \right|_{H_{C}^{-1} ( 0 )}
=&
- p_{i} \frac{ \partial H_{C} }{ \partial p_{i} } \frac{\partial}{ \partial \gamma} 
\\
& + \left( 
 \frac{ \partial H_{C} }{ \partial p_{i} } \frac{ \partial }{ \partial q^{i} }
- \frac{ \partial H_{C} }{ \partial q^{i} } \frac{ \partial }{ \partial p_{i} }
\right)
\\
& \hspace{-8mm} + 
\left( \Delta V - \mathbb{E}_{\pi_{H_{\beta}}} \! \left[ \Delta V \right] \right)
\beta \left( 1 - \beta \right)
p_{i} \frac{ \partial }{ \partial p_{i} }.
\end{align*}
Because the contact Hamiltonian flow preserves the lift of the target distribution 
by construction, it features all of the properties we were lacking in simulated 
annealing and simulated tempering.  

The Reeb component generates dynamic updates to the temperature, avoiding 
the need for a pre-defined partition.  These updates are coherent and avoid the 
random exploration that can limit simulated tempering -- for example, when the 
disintegration is constructed from a Riemannian metric, $g$, 
\cite{GirolamiEtAl:2011, BetancourtEtAl:2014},
\begin{align}
T \! \left( q, p \right) 
&= 
A \cdot F \! \left( g^{-1}\! \left( p, p \right) \right)
\nonumber
\\
& \quad +
\frac{1}{2} \log \left| g \! \left( q \right) \right| + \mathrm{const},
\label{riemann_disint}
\end{align}
the updates are monotonic.

Between temperature updates, the symplectic and Liouville components maintain 
the equilibrium that simulated annealing lacks.  The Liouville component of the flow 
can also be thought of as a perfect thermostat, in comparison to the approximate 
thermostats, such as the Nos\'{e}--Hoover thermostat \cite{EvansEtAl:1985}, 
common to molecular dynamics.

\begin{figure}
\centering
\includegraphics[width=2.75in]{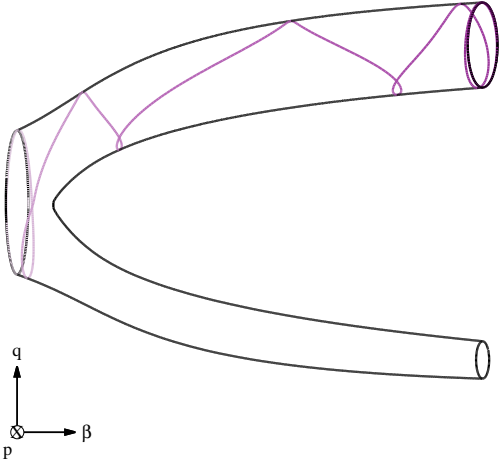}
\caption{Contact Hamiltonian flow generates trajectories that move through a 
continuum of Hamiltonian systems between the Hamiltonian system corresponding 
to the simple base distribution at $\beta = 0$ and the Hamiltonian system corresponding 
to the potentially complex target distribution at $\beta = 1$.  The trajectories 
corresponding to adiabatic processes dynamically adapt to maintain equilibrium at each 
temperature.}
\label{fig:free_level_set}
\end{figure}

In other words, adiabatic processes gives us the directed temperature exploration 
of simulated annealing, the equilibrium maintenance of simulated tempering, and 
a dynamic temperature partition that neither enjoy (Figure \ref{fig:free_level_set}).

An additional benefit of adiabatic processes is immediate recovery of the partition 
function $Z \! \left( \beta \right)$ at any time along the flow.  Because the contact 
Hamiltonian vanishes on the zero level set, the partition function is given at any 
point by
\begin{equation} \label{eqn:log_partition_estimator}
- \log Z \! \left( \beta \right)  = T + V_{B} + \beta \, \Delta V + H_{0},
\end{equation}
provided that the densities are all properly normalized.
Unlike most thermodynamic integration methods \cite{GelmanEtAl:1998},
this is an instantaneous result and does not require any quadrature.

\begin{figure}
\centering
\includegraphics[width=2.75in]{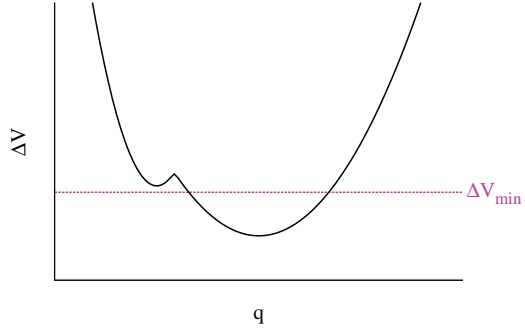}
\caption{Metastable equilibria occur when the Liousville component of the contact
Hamiltonian flow forces the momenta to zero too rapidly.  Consequently the flow relaxes into
a local minimum such as $\Delta V_{\mathrm{min}}$ and the temperature ceases
to update.}
\label{fig:metastable}
\end{figure}

\begin{figure}
\centering
\includegraphics[width=2.75in]{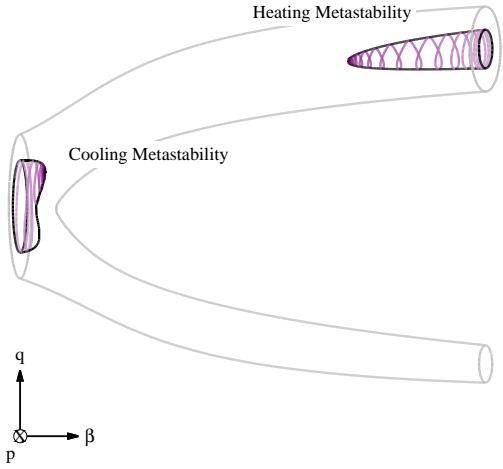}
\caption{A potential pathology of contact Hamiltonian flow are fixed points, or metastable
states.  Cooling metastabilities arise when a trajectory originating from $\beta = 0$ stalls,
while heating metastabilities arise when a trajectory originating from $\beta = 1$ stalls.
When a contact Hamiltonian flow suffers from metastabilities it is not an bijection
between the two temperatures and samples from one distribution are not necessarily
mapped to samples from the other.}
\label{fig:metastable_level_sets}
\end{figure}

Unfortunately, the contact Hamiltonian systems produced in this 
construction are not immune to fixed points.  For example, if we take a 
Riemannian disintegration \eqref{riemann_disint} then fixed points arise when 
the target parameters settle into a minimum of the effective potential energy, 
$V_{B} + \beta \, \Delta V$, the momenta fall to zero, and the flow along $\gamma$ 
stops (Figure \ref{fig:metastable}).  These fixed points correspond to metastable 
states in thermodynamics; \textit{cooling metastabilities} are accessed by flowing 
from $\beta = 0$ to $\beta = 1$ while \textit{heating metastabilities} are accessed 
by flowing from $\beta = 1$ to $\beta = 0$ (Figure \ref{fig:metastable_level_sets}).  

Metastable states obstruct the flow from being an bijection between $\beta = 0$ and 
$\beta = 1$.  For example, consider a \textit{cooling transition} from $\beta = 0$ to $\beta = 1$, 
or, recalling \eqref{eqn:logistic}, $\gamma = -\infty$ to $\gamma = \infty$.  Cooling metastabilities 
restrict the preimage of the flow and obstruct its surjectivity, while heating metastabilities restrict 
the image of the flow and obstruct its injectivity.

\subsection{Implementing Adiabatic Monte Carlo}

Once a disintegration has been chosen, in theory Adiabatic Monte Carlo proceeds 
similar to Hamiltonian Monte Carlo.  A sample from the base distribution,
\begin{equation*}
q_{i} \sim \pi,
\end{equation*}
is first lifted to the cotangent bundle by sampling from the local fiber,
\begin{equation*}
q_{i} \mapsto z_{i} = (q_{i}, p_{i}), \, p_{i}\sim \iota_{q_{i}}^{*} \xi,
\end{equation*}
and then to the contact manifold by setting $\beta = 0$.  The constant
$H_{0}$ is chosen such that the initial point falls on $H_{C}^{-1} \! \left( 0 \right)$
and the system is evolved backwards in time until $\beta = 1$ via a cooling transition,
\begin{equation*}
\left( z_{f}, \beta = 1\right) = \phi^{H_{C}}_{-t} \! \left( z_{i}, \beta = 0 \right),
\end{equation*}
and then projected back down to the sample space,
\begin{equation*}
q_{f} = \varpi \! \left( z_{f} \right).
\end{equation*}

Implementing this algorithm in practice, however, is significantly more complicated.
In addition to simulating the contact Hamiltonian flow, which requires not only an
accurate numeral integrator but also the accurate estimation of intermediate 
expectations and possibly temperature-dependent adaptation, we must also
overcome possible fixed points in the contact Hamiltonian flow.

\subsubsection{Simulating Contact Hamiltonian Flow}

Typically the contact Hamiltonian flow will not be solvable in practice and we 
must instead rely on a numerical approximation. Fortunately, contact Hamiltonian 
flow admits an accurate and robust numerical approximation in the same way that 
symplectic integrators approximate Hamiltonian flow.

Following the geometric construction of a symplectic integrator \cite{HairerEtAl:2006}, 
we can approximate the contact Hamiltonian flow by first splitting the contact Hamiltonian 
into three scalar functions,
\begin{align*}
H_{C} &= 
\underbrace{T}_{H_{1}} 
+ \underbrace{V_{B} }_{H_{2}} 
+ \underbrace{ \beta \Delta V + \log Z \! \left( \beta \right) }_{H_{3}} + H_{0}.
\end{align*}
This gives three vector fields along the contact structure,
\begin{align*}
\vec{X}_{H_{1}} &= 
\frac{ \partial T }{ \partial p_{i} } \left( \frac{ \partial }{ \partial q^{i} } 
- p_{i} \frac{ \partial }{ \partial \beta } \right)
- \frac{ \partial T }{ \partial q^{i} } \frac{ \partial }{ \partial p_{i} }
\\
\vec{X}_{H_{2}} &= - \frac{ \partial V_{B} }{ \partial q^{i} } \frac{ \partial }{ \partial p_{i} } \\
\vec{X}_{H_{3}} &= 
\left[ - \beta \frac{ \partial \Delta V }{ \partial q^{i} }
+ \left( \Delta V - \mathbb{E}_{\pi_{H_{\beta}}} \! \left[ \Delta V \right] \right) p_{i}  \right] \frac{ \partial }{ \partial p_{i} },
\end{align*}
and three corresponding contact Hamiltonian flows, 
$\phi_{t}^{H_{1}}$, $\phi_{t}^{H_{2}}$, and $\phi_{t}^{H_{3}}$.  
If the intermediate expectations 
$ \mathbb{E}_{\pi_{H_{\beta}}} \! \left[ \Delta V \right]$ are known 
then each of these flows can be solved immediately and their 
symmetric composition gives a reversible, second-order 
approximation to the exact flow (Algo \ref{algo:evolution}),
\begin{align*}
\phi_{\tau}^{H_{C}}
&=
\left( \phi_{\epsilon / 2}^{H_{1}} \circ
\phi_{\epsilon / 2}^{H_{2}} \circ
\phi_{\epsilon}^{H_{3}} \circ
\phi_{\epsilon / 2}^{H_{2}} \circ
\phi_{\epsilon / 2}^{H_{1}} \right)^{\tau / \epsilon} 
\\ 
& \quad 
+ \mathcal{O} \! \left( \epsilon^{2} \right).
\end{align*}
Because each component is a contact Hamiltonian flow, their composition is also a contact
Hamiltonian flow.  Consequently the numerical integration exactly preserves the contact volume 
form with only a small error in the contact Hamiltonian itself that can be controlled by manipulating
the integrator step size, $\epsilon$.

\begin{algorithm}
\caption{
Assuming that the expectations 
$ \mathbb{E}_{\pi_{H_{\beta}}} \! \left[ \Delta V \right]$ are known, a second-order
and reversible contact integrator is readily constructed by simulating flows from
component contact Hamiltonians.}
\label{algo:evolution}
\begin{algorithmic}

\STATE Initialize $\beta = 0$, $q \sim \pi_{\beta}$,  $p \sim \iota^{*}_{q} \xi$
\STATE $H_{0} \gets - \left( T \! \left( q, p \right) + V \! \left( q \right) \right)$
\STATE
\WHILE{ $\beta < 1$ }
  \STATE $\beta \gets \beta - \left( - \epsilon / 2 \right) p \cdot \partial T / \partial p $
  \hfill {\color{comment} $\phi^{H_{1}}_{-\epsilon / 2}$ }
  \STATE $q \gets q + \left( - \epsilon / 2 \right) \quad\,\, \partial T / \partial p $
  \STATE
  \STATE $p \gets p - \left( - \epsilon / 2 \right) \partial V_{B} / \partial q $
  \hfill {\color{comment}  $\phi^{H_{2}}_{-\epsilon / 2}$ }
  \STATE $p \gets p - \left( - \epsilon \right)  \beta \, \partial \Delta V / \partial q $
  \hfill {\color{comment} $\phi^{H_{3}}_{-\epsilon} \;\,$ }
  \STATE $ \quad\quad\quad\;\;
  + \left( - \epsilon \right) 
  \left( \Delta V -  \mathbb{E}_{\pi_{H_{\beta}}} \! \left[ \Delta V \right] \right) p $
  \STATE
  \STATE $p \gets p - \left( - \epsilon / 2 \right) \partial V_{B} / \partial q $
  \hfill {\color{comment}  $\phi^{H_{2}}_{-\epsilon / 2}$ }
  \STATE $\beta \gets \beta - \left( - \epsilon / 2 \right) p \cdot \partial T / \partial p $
  \hfill {\color{comment} $\phi^{H_{1}}_{-\epsilon / 2}$ }
  \STATE $q \gets q + \left( - \epsilon / 2 \right) \quad\,\, \partial T / \partial p $
\ENDWHILE
\STATE
\STATE {\color{comment} Proper normalization of $\Delta V$ required }
\STATE $- \log Z \! \left( \beta \right) = 
T \! \left( q, p \right) + V \! \left( q \right) + \beta \Delta V \! \left( q \right) + H_{0}$

\end{algorithmic}
\end{algorithm}

In order to exactly compensate for the error in the approximate integration of the contact
Hamiltonian flow we can appeal to the same Metropolis acceptance procedure often
used in Hamiltonian Monte Carlo, although with some slight modifications.  If there are
no metastabilties, for example, then a proposal targeting $\pi_{H_{\beta = 1}}$ can
be constructed by composing a heating transition with a momentum resampling and 
finally a cooling transition.  Given an initial state, $\left( q_{i}, p_{i} \right)$,
the final state $\left( q_{f}, p_{f} \right)$ can then be accepted with probability
\begin{align*}
p \! \left( \mathrm{accept} \right) &= 
\\
& \hspace{-5mm}
\min \! \left[ 1, \exp \! \left( 
H_{\beta = 1} \! \left( p_{i}, q_{i} \right) - H_{\beta = 1} \! \left( p_{f}, q_{f} \right)
\right) \right].
\end{align*}

Care must be taken with such a Metropolis correction, however, when the contact
Hamiltonian flow is subject to metastabilities and hence is not a proper bijection.

\subsubsection{Estimating Expectations}

Of course in any practical problem the expectations 
$\mathbb{E}_{\pi_{H_{\beta}}} \! \left[ \Delta V \right]$ will not be known a priori and 
we must instead estimate them online.  An immediate strategy is to use Hamiltonian 
Monte Carlo initialized at the current state, which should already be in local equilibrium.  

Running Hamiltonian Monte Carlo at each step of the contact integrator can quickly
become computationally limiting, and ensuring exact reversibility of the resulting
contact trajectories is a delicate problem.  A more robust approach is to run an ensemble 
of initial trajectories that estimate the expectations at each step.  These
intermediate expectations can be smoothed with a nonparametric estimator, such as a 
Gaussian process, and then used to implement accurate, fast, and exactly reversible 
trajectories.

When targeting multimodal distributions we have to be more careful still as each
initial trajectory will be able to estimate the expectations with respect to only the
local mode.  In order to construct an accurate global expectation we have to
weight the local expectations by the local partition functions, which are conveniently
provided by the contact Hamiltonian flow at no additional cost.

Finally, if we want to use the partition function then we must consider how the 
error in any such estimation scheme propagates to deviations in the contact
Hamiltonian.

\subsubsection{Adapting to Temperature-Dependent Curvature}

One of the powerful features of Hamiltonian Monte Carlo is that the disintegration
can be tuned to optimize the performance of a symplectic integrator in a certain
coordinate system.  

For example, Euclidean Hamiltonian Monte Carlo utilizes a Gaussian disintegration 
given by 
\begin{equation} \label{euclidean_distint}
T = \frac{1}{2} M^{-1} \! \left(p, p \right) + \frac{1}{2} \log \left| M \right|;
\end{equation}
when the inverse Euclidean Metric, $M^{-1}$, is aligned with the global
covariance of the coordinates, symplectic integrators can be run with
larger step sizes, lower costs, and fewer pathologies.  Similarly, Riemannian
Hamiltonian Monte Carlo utilizes a position-dependent metric aligned with
the local covariance of the target distribution.

Tuning the kinetic energy in Adiabatic Monte Carlo is more subtle given that
the curvature of $\pi_{\beta}$ can vary sharply with $\beta$.  In practice
this may require a temperature-dependent disintegration and a resultantly
more complicated contact Hamiltonian flow.  One of the advantages of a
Riemannian Hamiltonian Monte Carlo with the SoftAbs metric 
\cite{Betancourt:2013b} is that an optimal temperature-dependent tuning 
is given implicitly by the metric itself.

\subsubsection{Overcoming Metastabilities}

As noted above, the contact Hamiltonian flow may exhibit fixed points which
manifest as metastable states obstructing and obstruct the flow from being
a bijection.

Fortunately both cooling and heating metastabilities can be overcome by 
simply resampling the momentum often enough -- resampling near a cooling 
metastability kicks the flow out of the local minimum ensuring subjectivity while 
resampling after a heating metastability allows the flow to access all final states 
and recovers injectivity.  Moreover, if $H_{0}$ is incremented with the difference 
in kinetic energies before and after the resampling,
\begin{equation*}
H_{0} \rightarrow H_{0} + T_{\mathrm{before}} - T_{\mathrm{after}},
\end{equation*}
then the partition function can still be recovered from \eqref{eqn:log_partition_estimator}.

The implementation challenge is in exactly when to interrupt the contact 
Hamiltonian flow with a momentum resampling.  Because the temperature evolution 
slows as the flow approaches a cooling metastability, resampling the momentum 
at uniform time intervals is sufficient to avoid the metastability itself.  Recovering
from heating metastabilities, however, is more challenging because the metastability
has no immediate impact on the flow.  Instead we can only assume the presence of
heating metastabilities and resample often as the flow approaches $\beta = 1$.
Additionally, if we want to apply a Metropolis correction then any resampling
scheme must also be reversible.

\section{Beta-Binomial Example}

\begin{figure*}
\centering
\subfigure[]{ \label{fig:density_vs_temp} \includegraphics[width=1.91in]{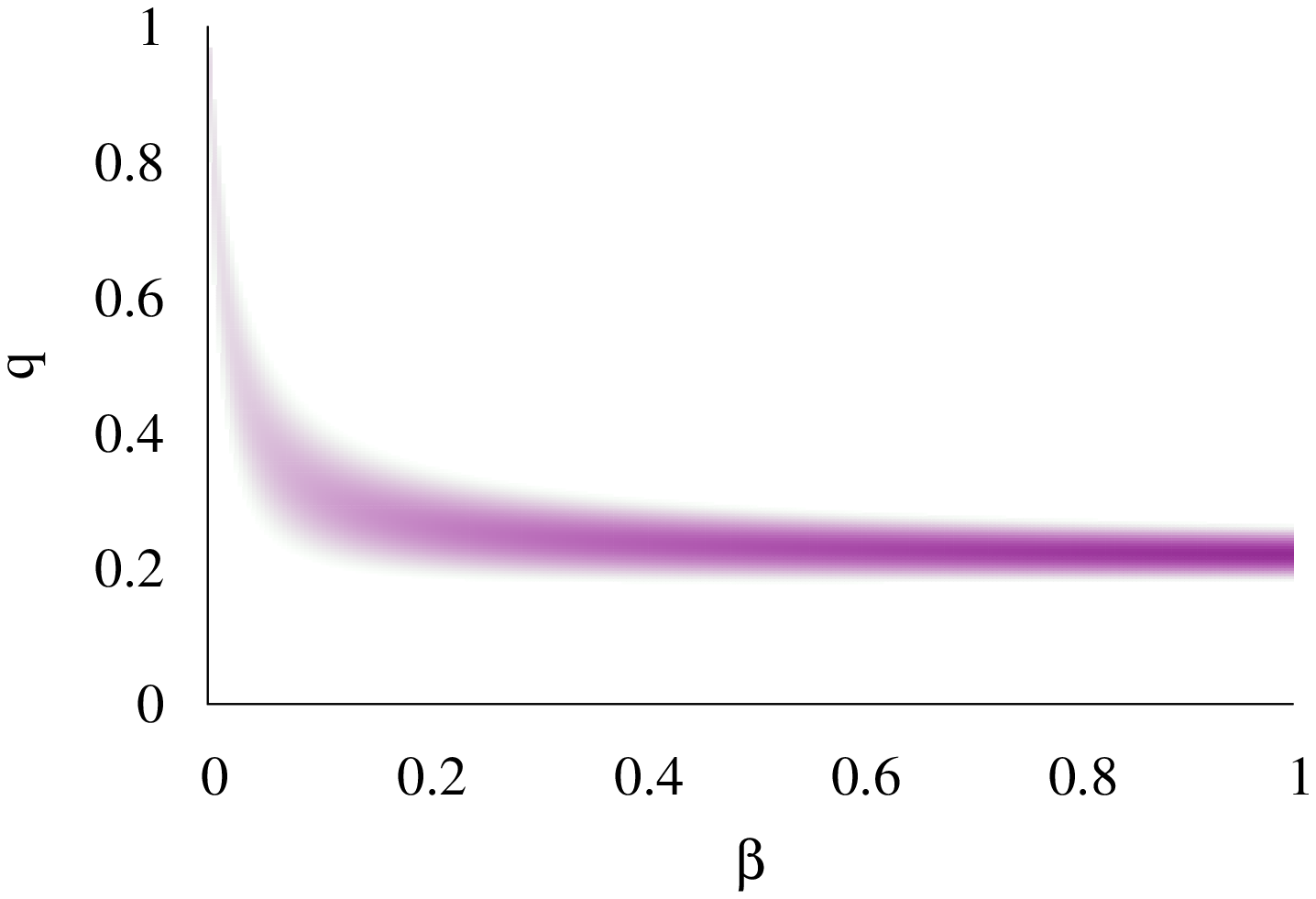}}
\subfigure[]{ \label{fig:density_at_boundary} \includegraphics[width=1.91in]{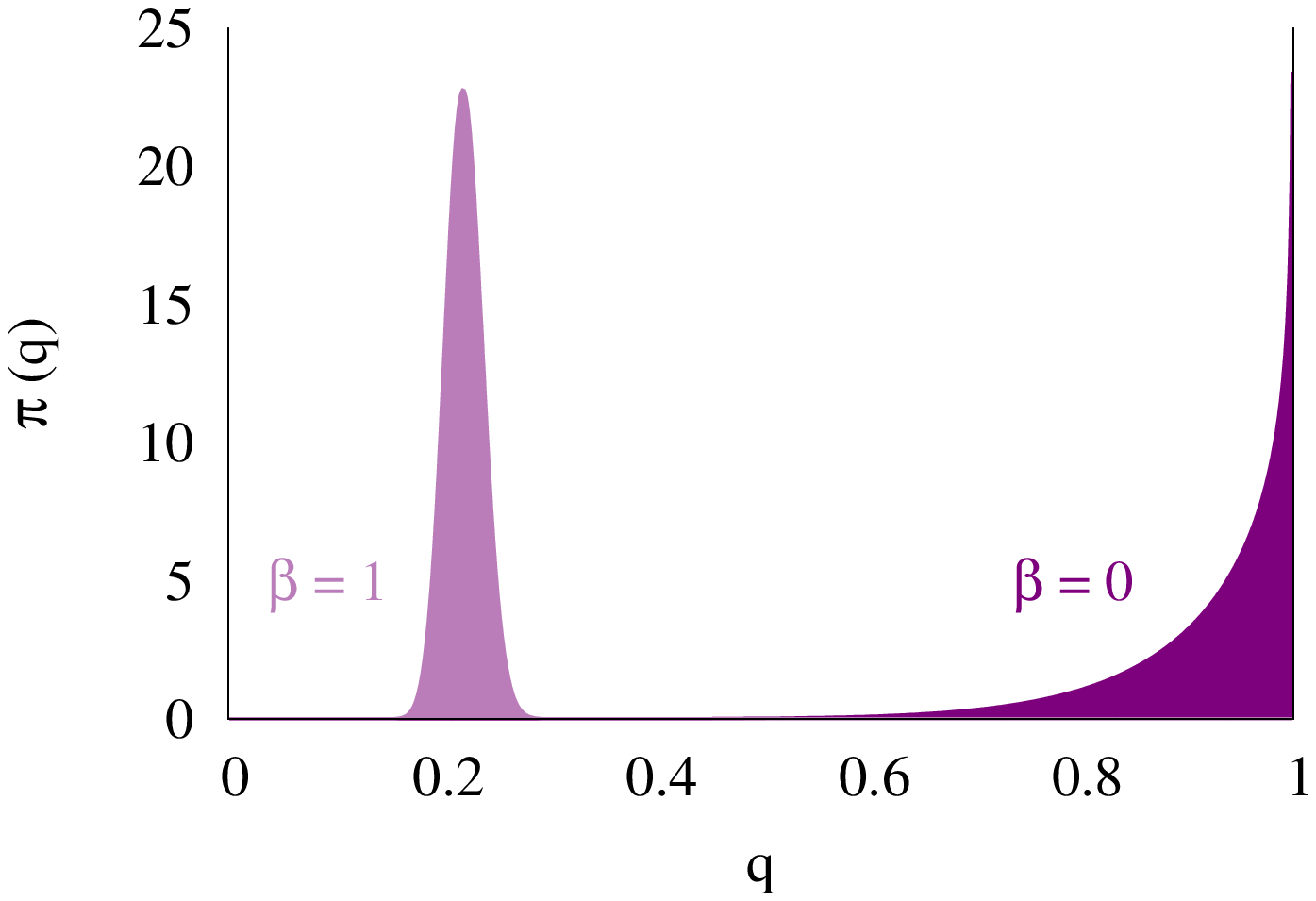}}
\subfigure[]{ \label{fig:partition} \includegraphics[width=1.91in]{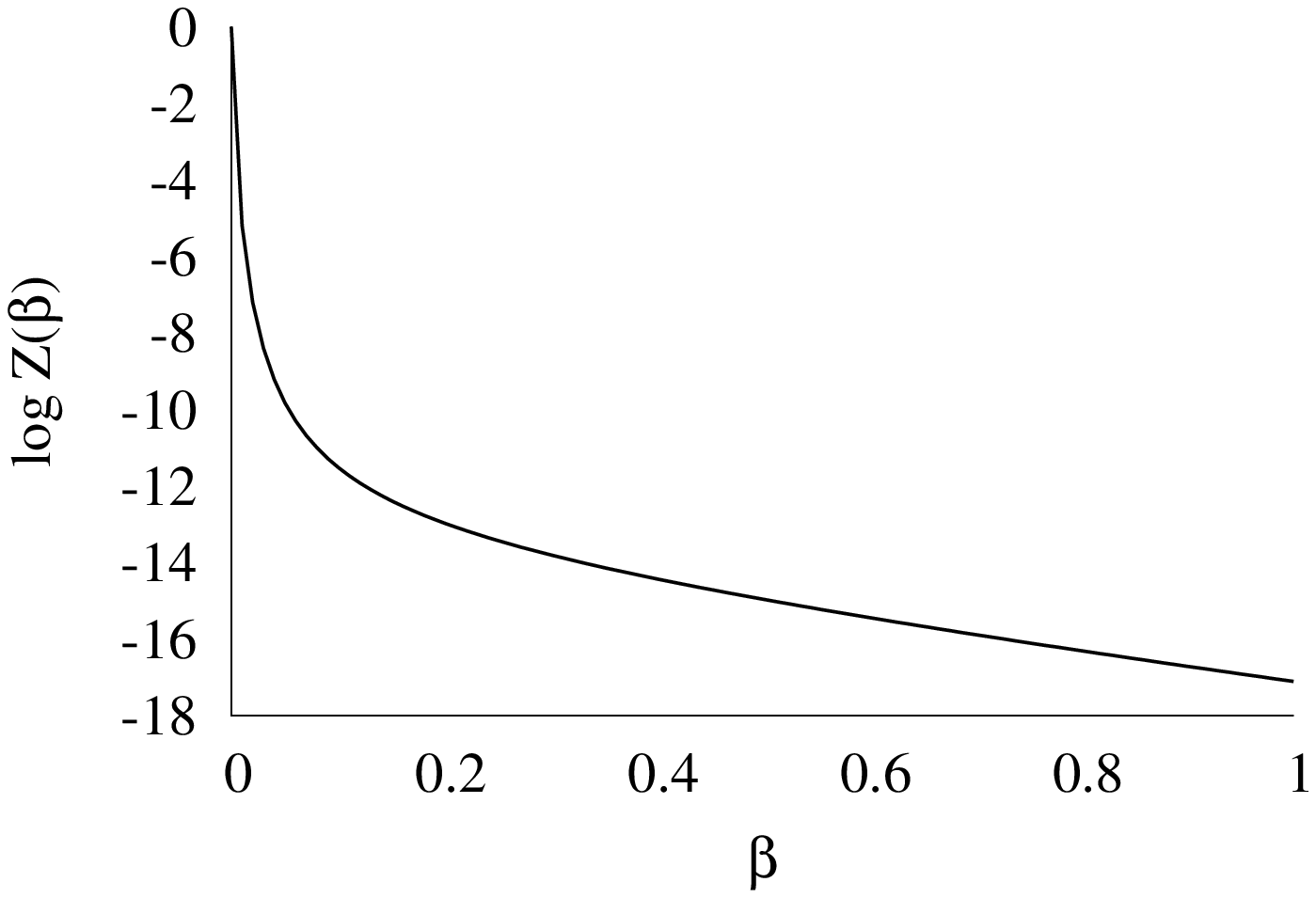}}
\caption{Emulative of the difficult models encountered in practice, the target 
and base distributions in the experiments have little overlap which exposes the 
weakness of simulated annealing and simulated tempering.  Here (a, b) the 
probability mass rapidly slides from one boundary towards the other as the inverse 
temperature, $\beta$ increases from 0 to 1.  Consequently, (c) the partition function 
is also extremely sensitive to the inverse temperature.}
\end{figure*}

In order to demonstrate the power of adiabatic processes without concerning
ourselves with metstabilties, let us a examine a simple one-dimensional and 
univariate example.  In particular, consider the Beta distribution taking the role 
of both the target and base distributions with a Binomial density between them,
\begin{align*}
\pi_{\beta} 
&\propto
\left( \mathrm{Bi} \! \left(k | n \right) \right)^{\beta} \mathrm{Be} \! \left( a, b \right)
\\
&= \mathrm{Be} \! \left( \beta \, k + a, \beta \left( n - k \right) + b \right).
\end{align*}
In this case we can analytically compute both the partition function,
\begin{align*}
Z \! \left( \beta \right) 
&=
\left[ \frac{ \Gamma \! \left( n + 1 \right) }
{ \Gamma \! \left( k  + 1 \right) \Gamma \! \left( n - k + 1 \right) } \right]^{\beta}
\frac{ \Gamma \! \left( a + b \right) }{ \Gamma \! \left( a \right) \Gamma \! \left( b \right) } \\
&=
\quad \times
\frac{ 
\Gamma \! \left( \beta \, k + a \right) 
\Gamma \! \left( \beta \left( n - k \right) + b \right) }
{ \Gamma \! \left( \beta \, n + a + b \right ) },
\end{align*}
and its derivative,
\begin{align*}
\frac{1}{ Z \! \left( \beta \right) } \frac{ \partial Z \! \left( \beta \right) }{ \partial \beta } 
&= \;\;\,
\log \! \left( \frac{ \Gamma \! \left( n + 1 \right) }
{ \Gamma \! \left( k  + 1 \right) \Gamma \! \left( n - k + 1 \right) } \right)
\\
& \quad + k \, \psi  \! \left( \beta \, k + a \right)
\\
& \quad + \left( n - k \right) \psi \! \left( \beta \left( n - k \right) + b \right) 
\\
& \quad - n \, \psi \! \left( \beta \, n + a + b \right) .
\end{align*}
In the following I take $a = 9$, $b = 0.75$, $k = 115$, and $n = 550$ such that
the target and base distributions have only small overlap (Figure \ref{fig:density_vs_temp},
\ref{fig:density_at_boundary}) and a rapidly changing partition function (Figure \ref{fig:partition}).

Given these analytic results we can readily investigate the performance of simulated 
annealing, simulated tempering, and then Adiabatic Monte Carlo.  Note that with
the preponderance of analytic results in this example, both simulated annealing and simulated 
tempering can be tuned to achieve reasonable performance.  Our goal here is not to 
demonstrate that the existing algorithms fail in this simple case but rather to exemplify 
the kinds of pathologies that become unavoidable when targeting complex distributions 
in high dimensions.

\subsection{Simulated Annealing}

Here I implemented simulated annealing with two Random Walk Metropolis \cite{RobertEtAl:1999} 
transitions in between temperature updates.  At each temperature I tuned the proposal scale to achieve
the optimal acceptance probability for a one-dimensional target distribution \cite{RobertsEtAl:1997}.  

I ran simulated annealing three times, each with a different partition of the inverse temperature: 
a coarse partition consisting of 25 evenly spaced intervals, a fine partition consisting of 100 evenly 
spaced intervals, and an optimally-tuned partition consisting of 25 intervals such that the 
Kullback-Leibler divergence between each partition is constant.

\begin{figure*}
\centering
\subfigure[]{ \label{fig:anneal_coarse} \includegraphics[width=1.91in]{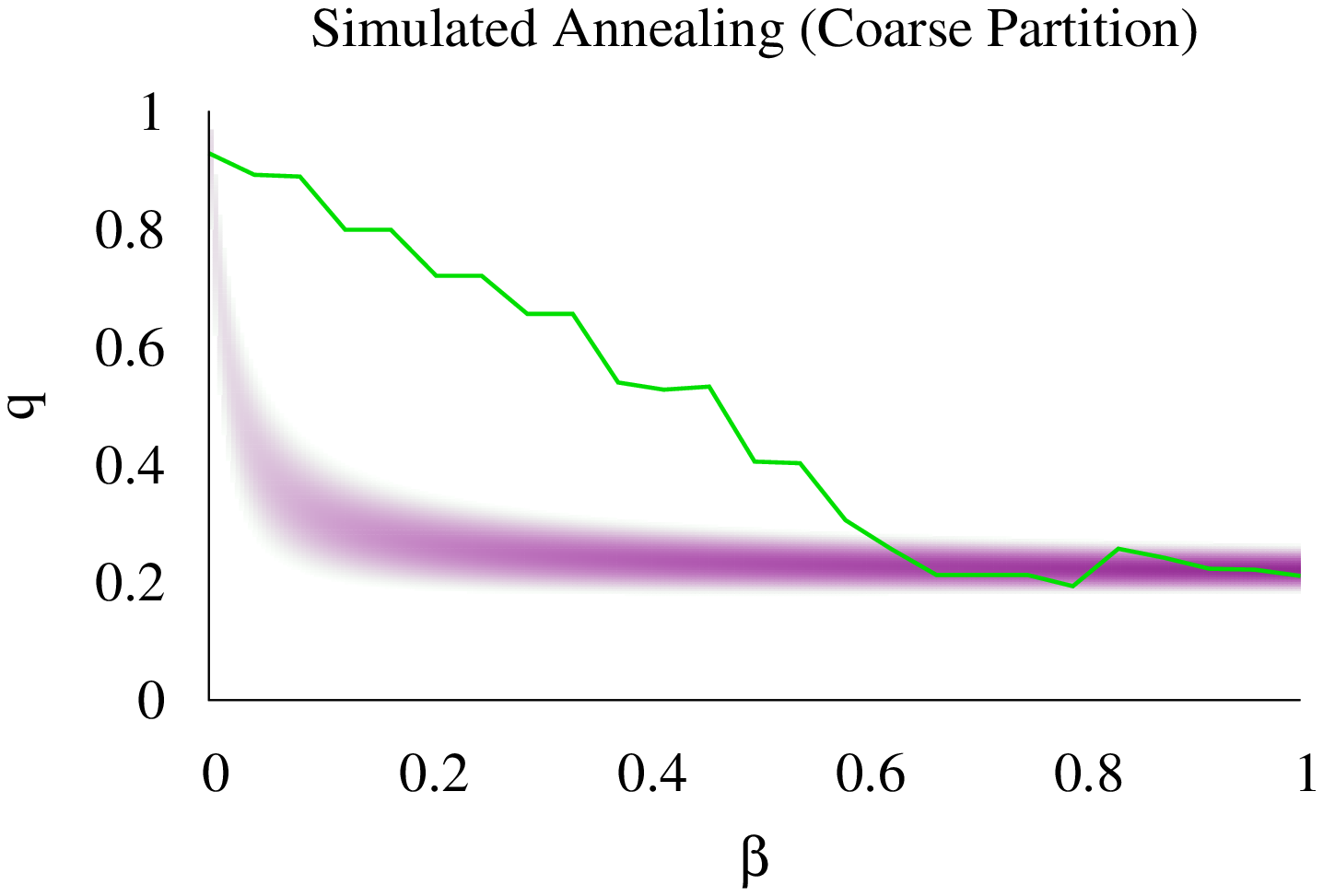}}
\subfigure[]{ \label{fig:anneal_fine} \includegraphics[width=1.91in]{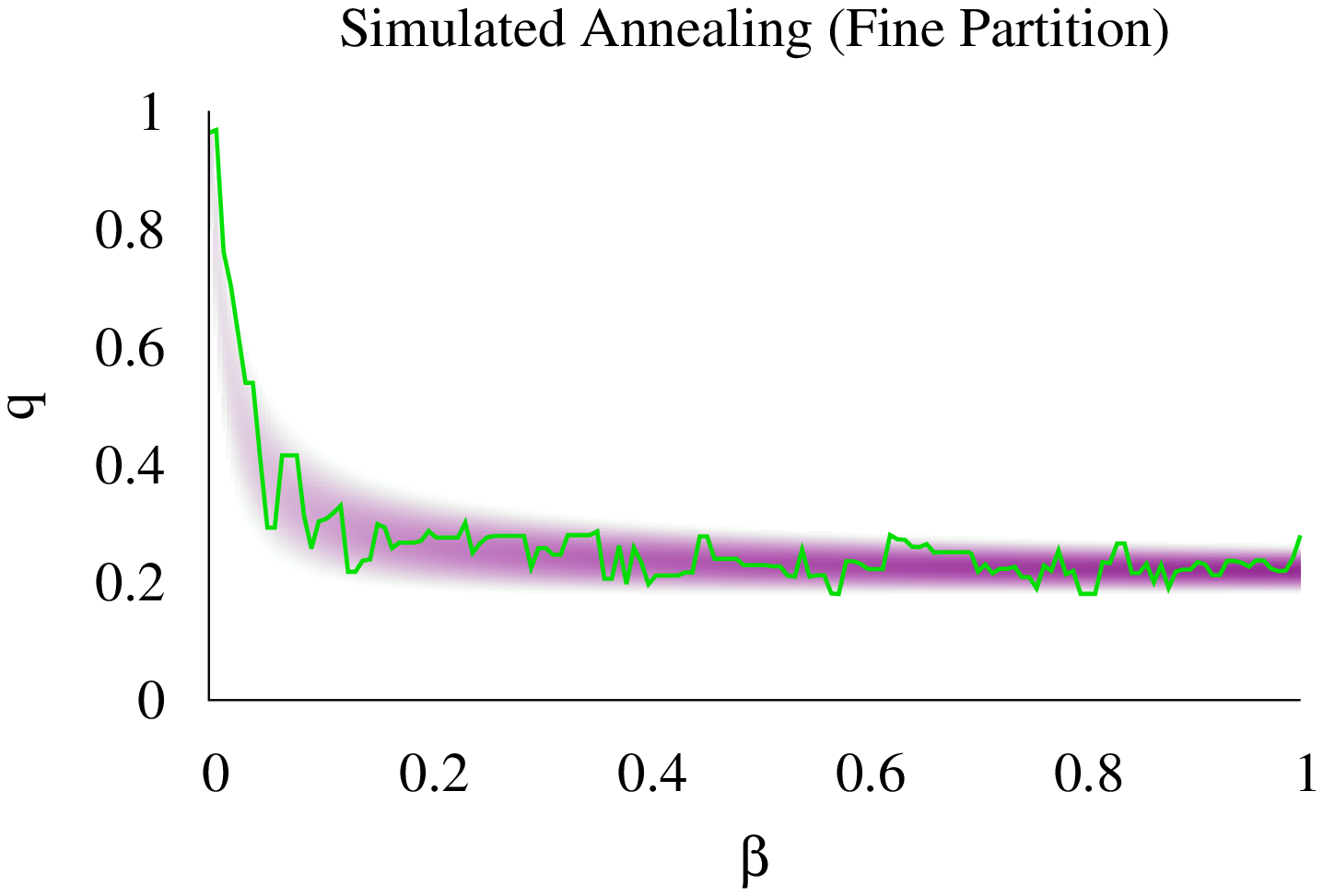}}
\subfigure[]{ \label{fig:anneal_tuned} \includegraphics[width=1.91in]{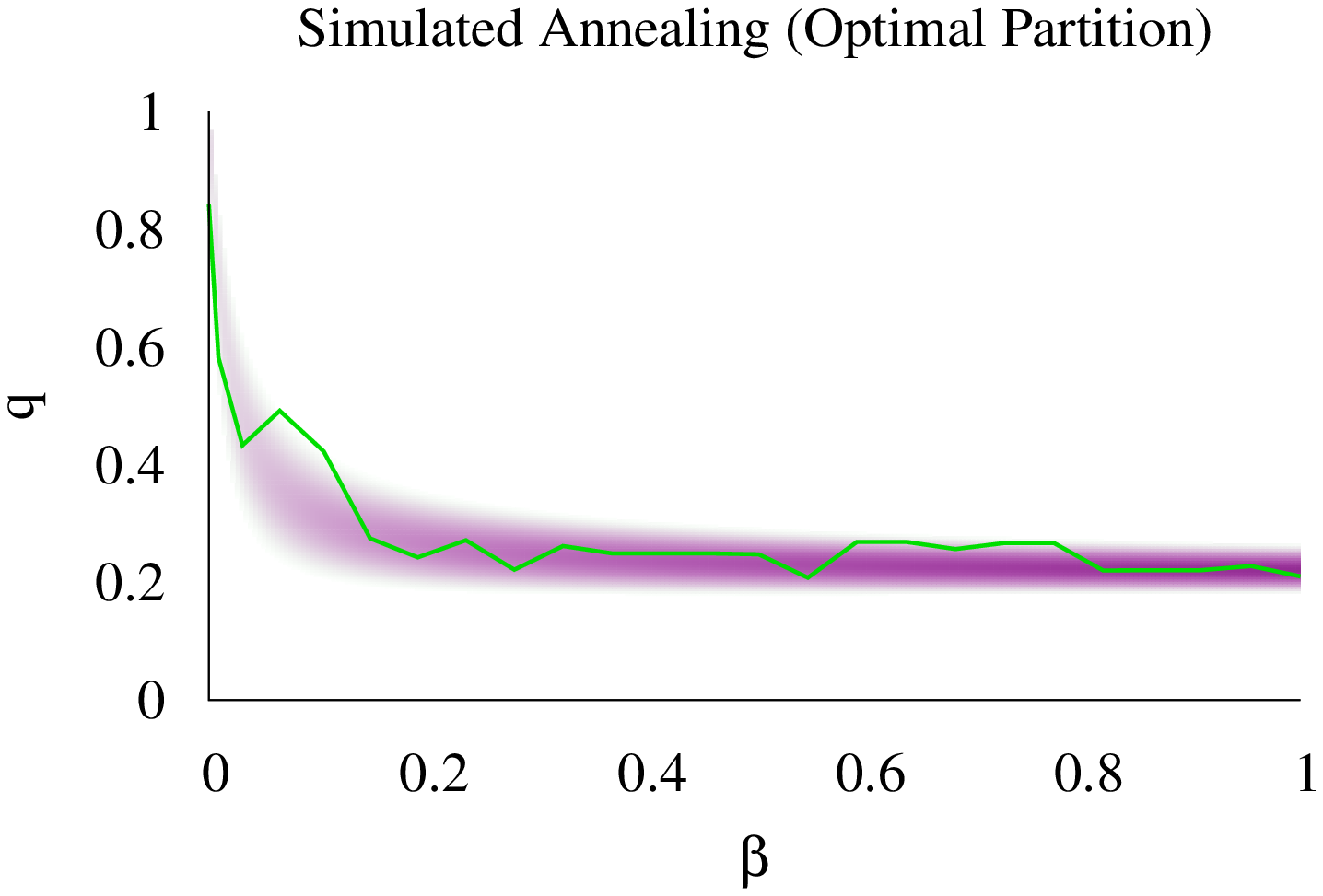}}
\caption{Even for this one-dimensional target distribution simulated annealing can
(a) rapidly fall out of equilibrium for coarse partitions.  Only with (b) very fine and (c) 
optimally-tuned partitions can equilibrium be maintained.}
\end{figure*}

The coarse partition is not well-tuned to the local variations in $\pi_{\beta}$; the
state rapidly falls out of equilibrium and then converges only well after the target
distribution stops changing with temperature (Figure \ref{fig:anneal_coarse}).  Only
with much smaller (Figure \ref{fig:anneal_fine}) and optimally-tuned (Figure \ref{fig:anneal_tuned})
partitions does the state remain in equilibrium throughout the entire transition.

The biggest weakness of simulated annealing is not so much that the state can fall out of 
equilibrium but rather that falling out of equilibrium can be extremely difficult to diagnose in practice.
As the target distribution becomes more complex, especially as it grows in dimensionality,
the potential for falling out of equilibrium and not re-converging becomes greater and greater.
Consequently, simulated annealing is not a particularly robust choice for statistical applications.

\subsection{Simulated Tempering}

As above, I implemented simulated tempering three times, using Random Walk Metropolis optimally 
tuned to the coarse, fine, and tuned partitions.  After 25 warmup transitions at $\beta = 0$ the chain 
evolves by jumping between neighboring temperatures in the partition.

\begin{figure*}
\centering
\subfigure[]{ \label{fig:temper_coarse} \includegraphics[width=1.91in]{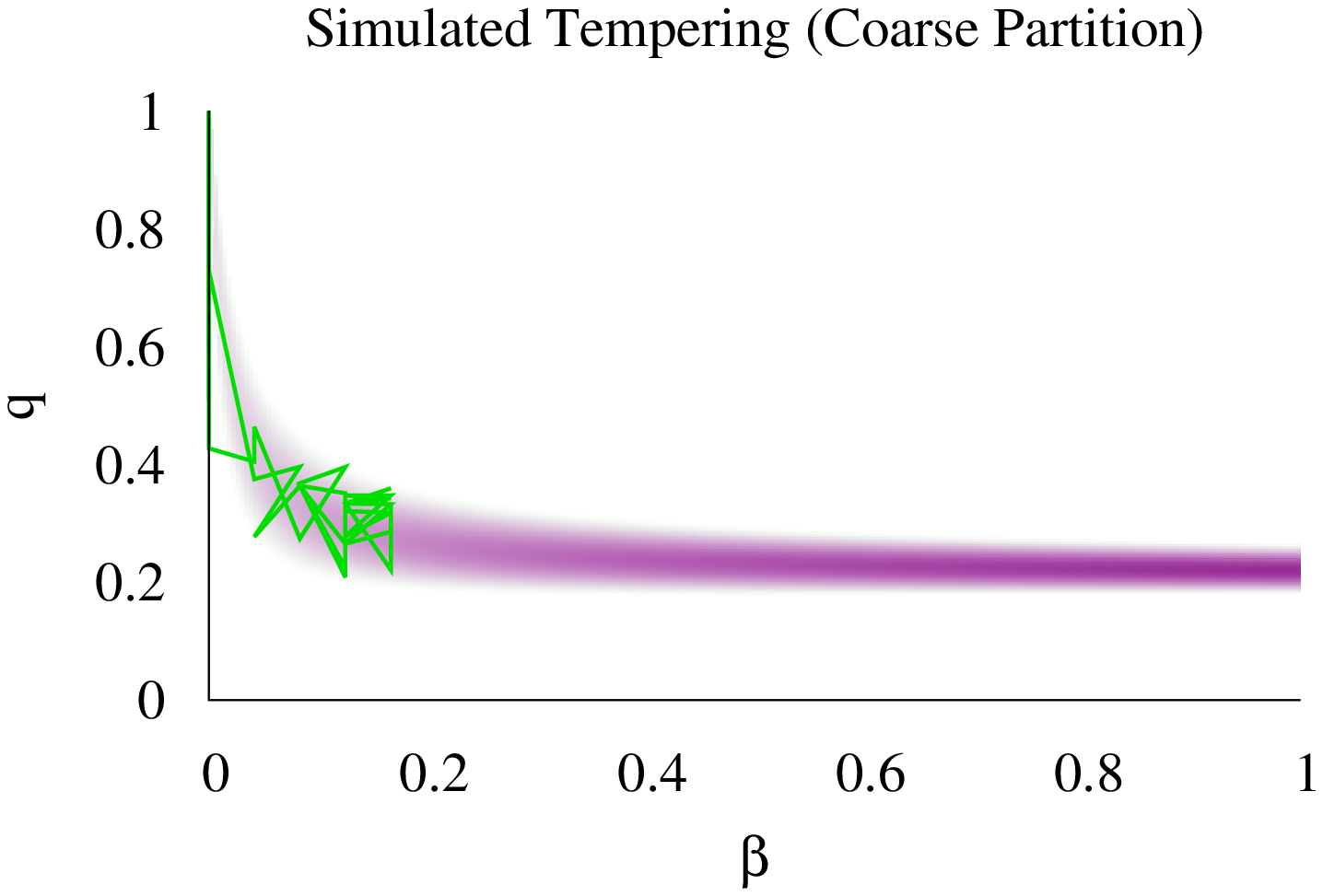}}
\subfigure[]{ \label{fig:temper_fine} \includegraphics[width=1.91in]{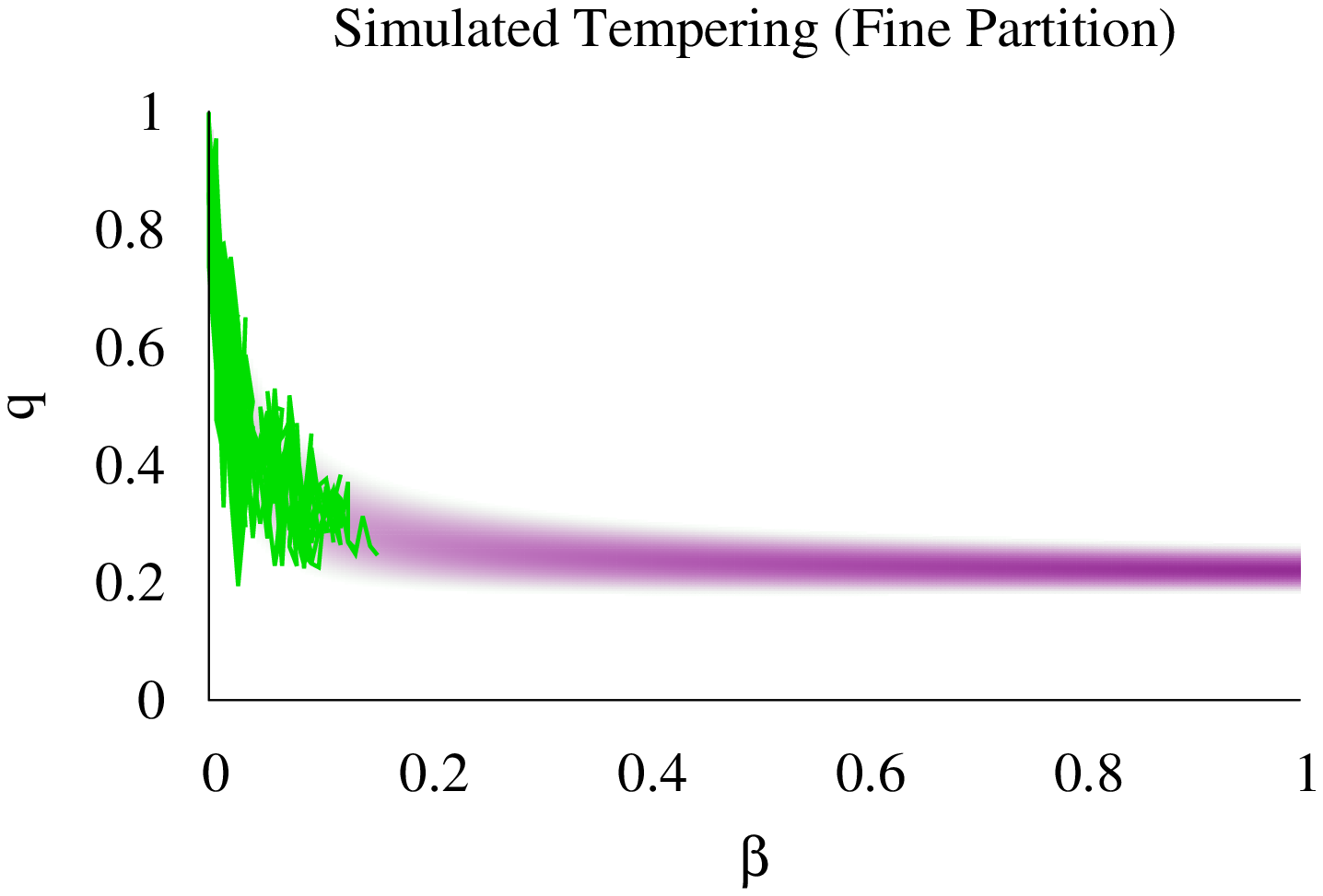}}
\subfigure[]{ \label{fig:temper_tuned} \includegraphics[width=1.91in]{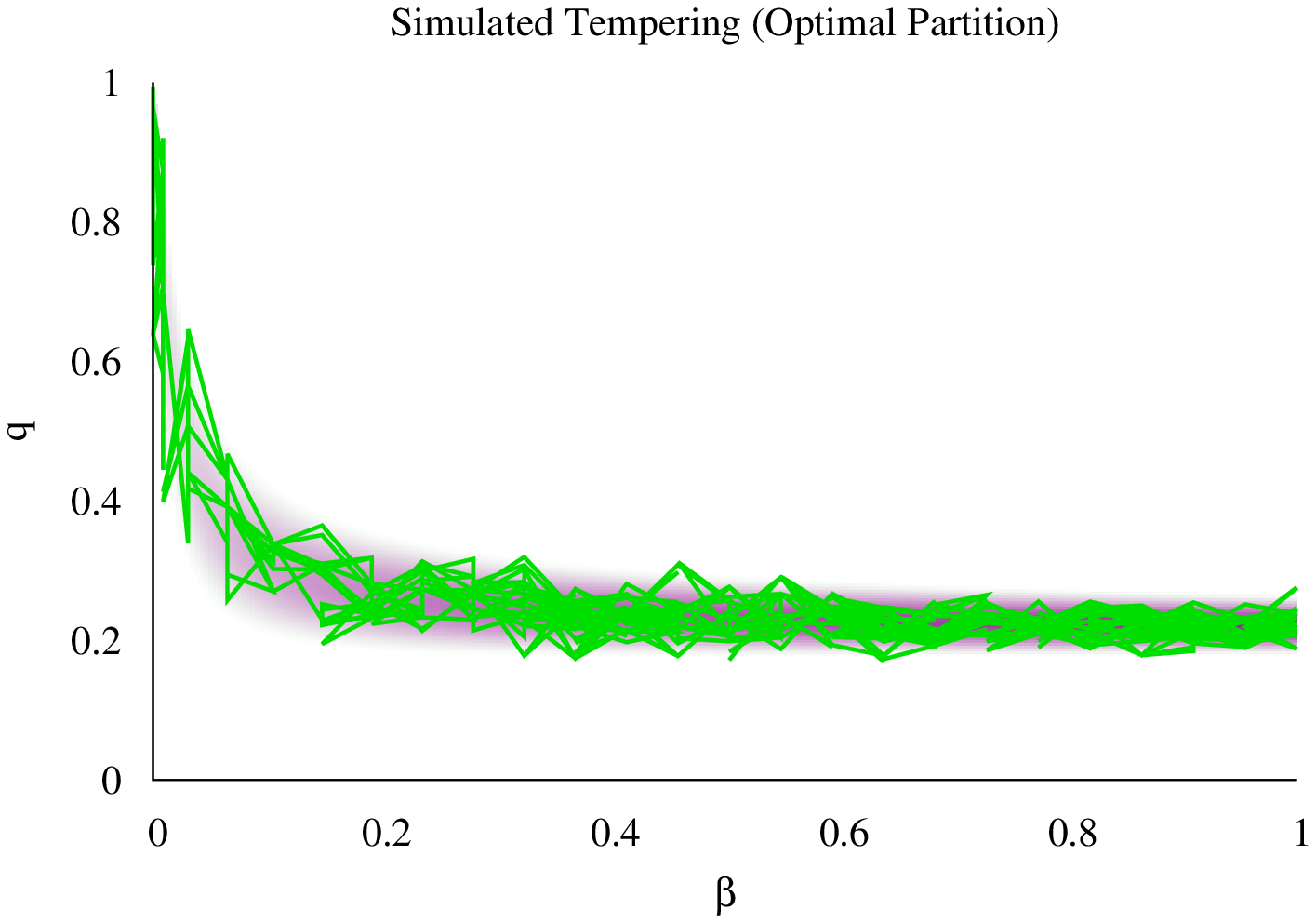}}
\caption{Although simulated tempering maintains the equilibrium that simulated annealing
is prone to losing, the efficacy of its temperature exploration depends critically on the configuration 
of the partition.  Both (a) the course partition and (b) the fine partition suffer from amplified random 
walk behavior.  (c) Only the tuned partition admits reasonably efficient exploration.}
\end{figure*}

In all three cases simulated tempering is able to maintain equilibrium as expected.  When 
using the coarse (Figure \ref{fig:temper_coarse}) and fine (Figure \ref{fig:temper_fine}) partitions,
however, the active state explores inefficiently never reaches $\beta = 1$.  Only with the tuned partition 
can information propagate between $\beta = 0$ and $\beta = 1$ in a reasonable amount of time 
(Figure \ref{fig:temper_tuned}).

More complex transitions between temperatures offer some hope of improving the inefficient 
exploration but in practice they are difficult to tune, especially when considering the high 
dimensional target distributions of interest.

\subsection{Adiabatic Monte Carlo}

I implemented Adiabatic Monte Carlo with the Gaussian Euclidean disintegration 
\eqref{euclidean_distint} and the resulting integrator as described in Algorithm 
\ref{algo:evolution}.  The expectation $ \mathbb{E}_{\pi_{H_{\beta}}} \! \left[ \Delta V \right] $ 
was estimated at each temperature using Hamiltonian Monte Carlo seeded at the current 
position of the chain.  For both the contact Hamiltonian flow and the intermediate 
Hamiltonian Monte Carlo runs the step size was set to $\epsilon = 0.01$, 
and the integration time for Hamiltonian Monte Carlo was randomly sampled as 
$\tau \sim U \left[ 0, 2 \varpi \right)$.

\begin{figure}
\centering
\subfigure[]{ \label{fig:thermo} \includegraphics[width=2.7in]{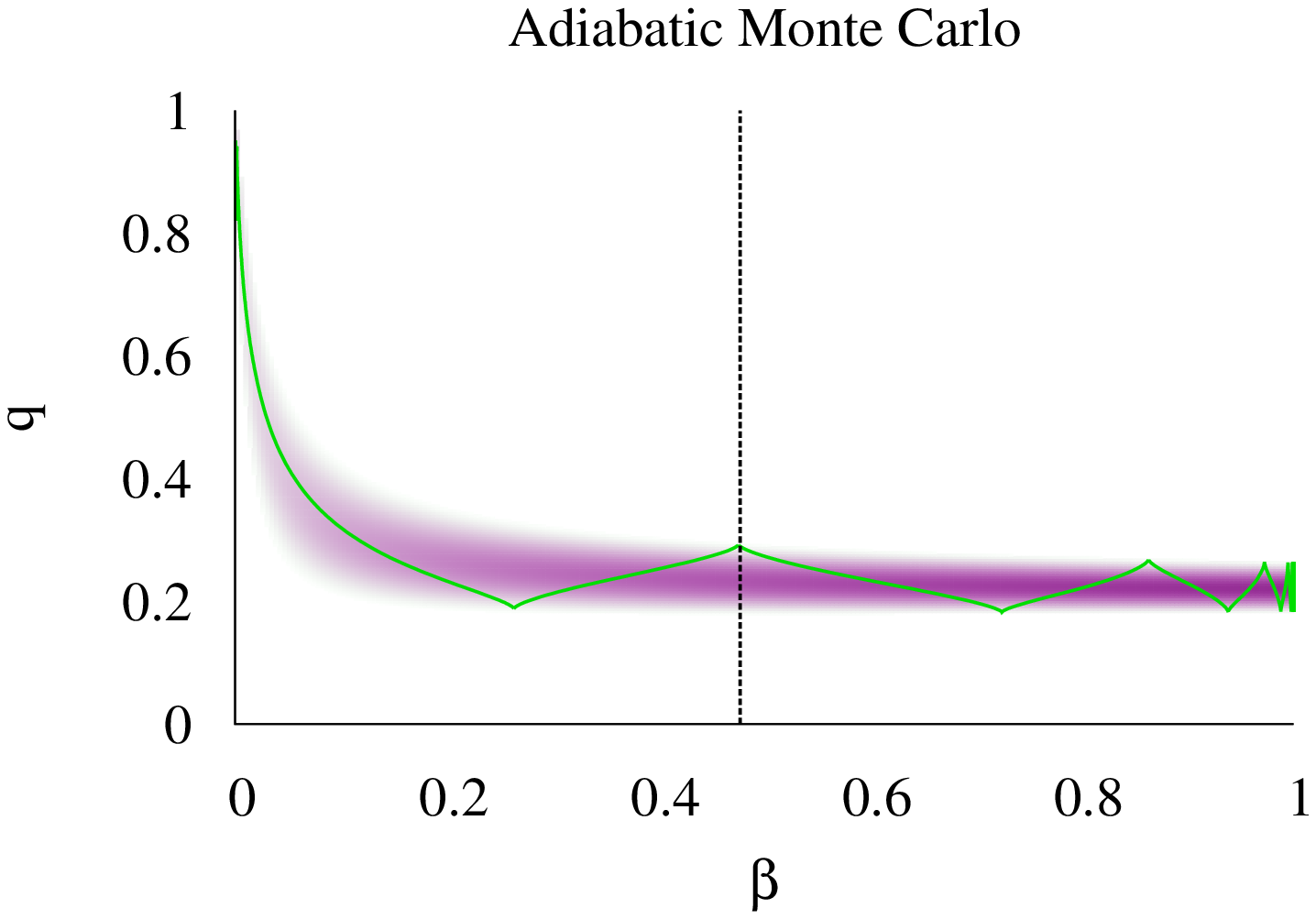}}
\subfigure[]{ \label{fig:thermo_delta_temp} \includegraphics[width=2.7in]{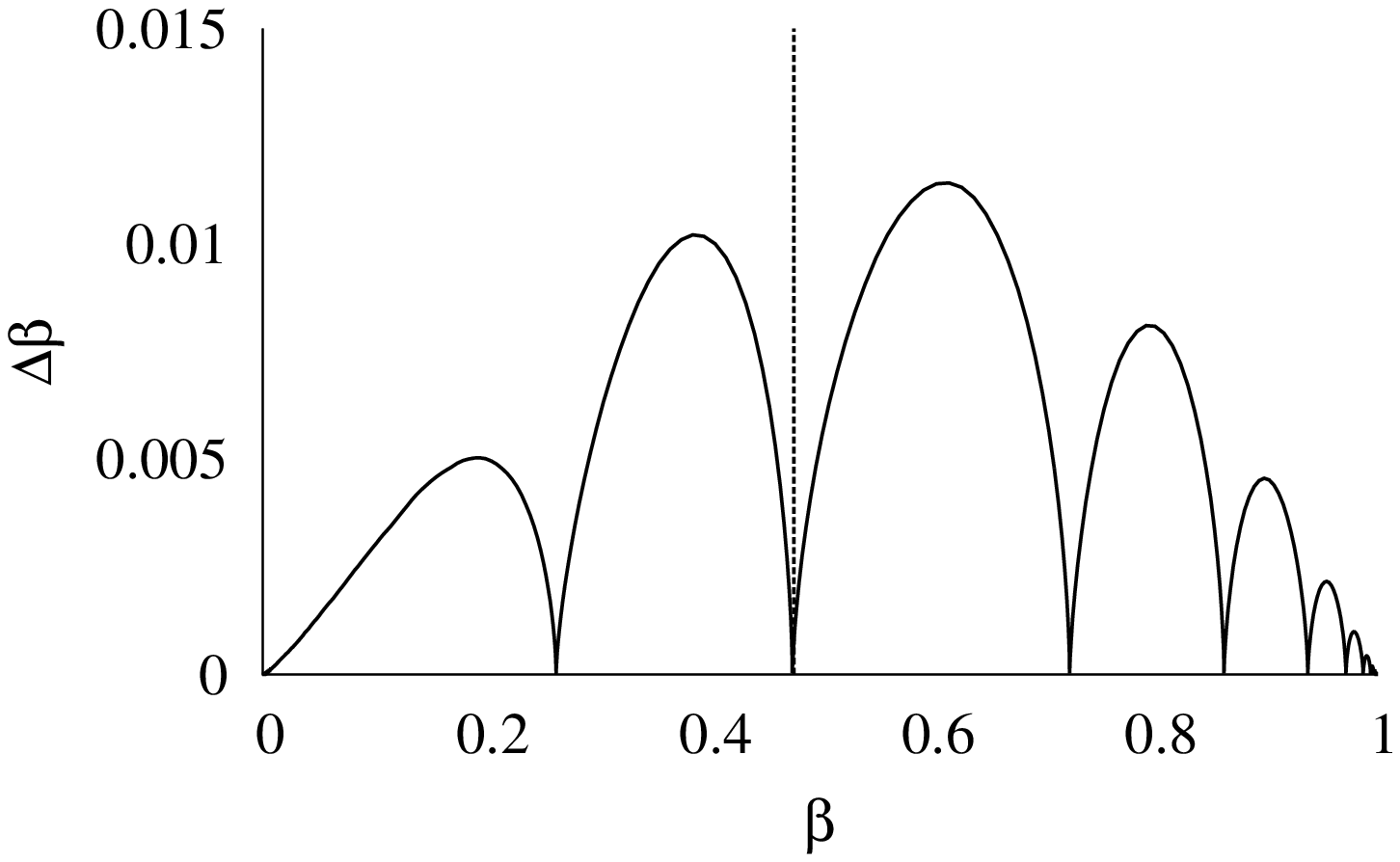}}
\caption{(a) Adiabatic Monte Carlo utilizes a contact Hamiltonian flow to transition between
temperatures while maintaining equilibrium.  (b) Because the temperature is a dynamic component
of the flow, the evolution effectively determines an optimal temperature partition dynamically.  As
the trajectory moves away from the probability mass, for example at the dotted line, the temperature
evolution slows to give the trajectory time to return to equilibrium.
}
\end{figure}

\begin{figure}
\centering
\includegraphics[width=2.75in]{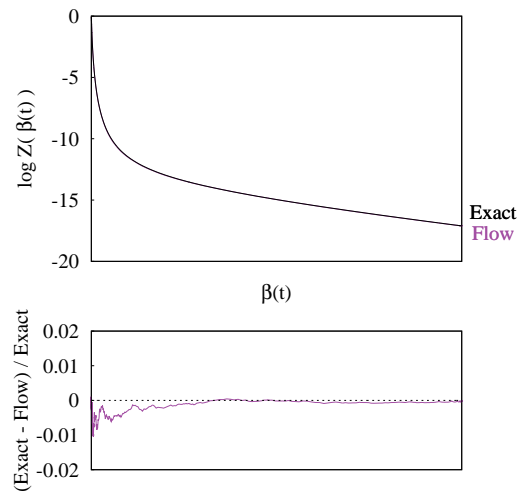}
\caption{A convenient byproduct of contact Hamiltonian flow is an accurate estimate of the
partition function at each inverse temperature, $\beta$.  The error in the estimate is too
small to be seen in the upper panel even as the true partition function varies across 8 orders 
of magnitude.}
\label{fig:thermo_partition_history} 
\end{figure}

Without requiring any temperature partition, Adiabatic Monte Carlo is able to maintain equilibrium 
while efficiently exploring all temperatures by effectively determining a partition dynamically 
(Figure \ref{fig:thermo}).  As desired, the temperature changes dynamically slow as the trajectory 
deviates away from equilibrium and increases only once the trajectory has returned to the bulk of the 
probability mass (Figure \ref{fig:thermo_delta_temp}).  Moreover, without any additional computation 
the trajectory also provides an accurate estimate of the partition function 
(Figure \ref{fig:thermo_partition_history}).

\section{Conclusion}

By leveraging the geometry of contact Hamiltonian systems, Adiabatic Monte Carlo admits
a uniquely powerful approach to exploring the complex and multimodal target distributions
that confound Markov Chain Monte Carlo algorithms.  Moreover, this foundational geometry
not only identifies potential pathologies, such as metastabilities, but also guides the construction 
of the implementations robust to those pathologies.  Algorithms incorporating this guidance
are currently under development with the ultimate goal an implementation in Stan \cite{Stan:2014}.

\section{Acknowledgements}

I thank Tarun Chitra, Andrew Gelman, Mark Girolami, Matt Johnson, and Stephan Mandt for thoughtful 
comments and Chris Wendl for illuminating contact geometries.  This work was  supported by EPSRC 
grant EP/J016934/1.

\bibliography{adiabatic_monte_carlo}

\begin{thebibliography}{10}

\bibitem{RobertEtAl:1999}
C.~P. Robert and G.~Casella,
\newblock {\em {M}onte {C}arlo Statistical Methods} (Springer New York, 1999).

\bibitem{BrooksEtAl:2011}
S.~Brooks, A.~Gelman, G.~L. Jones, and X.-L. Meng, editors,
\newblock {\em Handbook of {M}arkov {C}hain {M}onte {C}arlo} (CRC Press, New
  York, 2011).

\bibitem{DuaneEtAl:1987}
S.~Duane, A.~Kennedy, B.~J. Pendleton, and D.~Roweth,
\newblock Physics Letters B {\bf 195}, 216  (1987).

\bibitem{Neal:2011}
R.~Neal,
\newblock {MCMC} using {H}amiltonian dynamics,
\newblock in {\em Handbook of Markov Chain Monte Carlo}, edited by S.~Brooks,
  A.~Gelman, G.~L. Jones, and X.-L. Meng, CRC Press, New York, 2011.

\bibitem{BetancourtEtAl:2014}
M.~Betancourt, S.~Byrne, S.~Livingstone, and M.~Girolami,
\newblock ArXiv e-prints {\bf 1410.5110} (2014).

\bibitem{KirkpatrickEtAl:1983}
S.~Kirkpatrick {\em et~al.},
\newblock Science {\bf 220}, 671 (1983).

\bibitem{Cerny:1985}
V.~{\v{C}}ern{\o{y}},
\newblock Journal of Optimization Theory and Applications {\bf 45}, 41 (1985).

\bibitem{Neal:1993}
R.~Neal,
\newblock Department of Computer Science, University of Toronto Report No.
  CRG-TR-93-1, 1993 (unpublished).

\bibitem{MarinariEtAl:1992}
E.~Marinari and G.~Parisi,
\newblock Europhysics Letters {\bf 19}, 451 (1992).

\bibitem{Geiges:2008}
H.~Geiges,
\newblock {\em An Introduction to Contact Topology} (Cambridge Univ. Press,
  2008).

\bibitem{Lee:2013}
J.~M. Lee,
\newblock {\em Introduction to Smooth Manifolds} (Springer, 2013).

\bibitem{Mrugala:1978}
R.~Mruga{\l}a,
\newblock Reports on Mathematical Physics {\bf 14}, 419 (1978).

\bibitem{GirolamiEtAl:2011}
M.~Girolami and B.~Calderhead,
\newblock Journal of the Royal Statistical Society: Series B (Statistical
  Methodology) {\bf 73}, 123 (2011).

\bibitem{EvansEtAl:1985}
D.~J. Evans and B.~L. Holian,
\newblock The Journal of Chemical Physics {\bf 83}, 4069 (1985).

\bibitem{GelmanEtAl:1998}
A.~Gelman and X.-L. Meng,
\newblock Statistical science , 163 (1998).

\bibitem{HairerEtAl:2006}
E.~Hairer, C.~Lubich, and G.~Wanner,
\newblock {\em Geometric Numerical Integration: {S}tructure-Preserving
  Algorithms for Ordinary Differential Equations} (Springer, New York, 2006).

\bibitem{Betancourt:2013b}
M.~Betancourt,
\newblock A general metric for {R}iemannian {H}amiltonian {M}onte {C}arlo,
\newblock in {\em First International Conference on the Geometric Science of
  Information}, edited by F.~Nielsen and F.~Barbaresco, , Lecture Notes in
  Computer Science Vol. 8085, Springer, 2013.

\bibitem{RobertsEtAl:1997}
G.~O. Roberts {\em et~al.},
\newblock The annals of applied probability {\bf 7}, 110 (1997).

\bibitem{Stan:2014}
{Stan Development Team},
\newblock {S}tan: {A} {C}++ library for probability and sampling, version 2.5,
  2014.

\end{thebibliography}
\bibliographystyle{PhysRevStyle}

\end{document}